\documentstyle[aps,pra,psfig,multicol]{revtex}
\draft
\begin{document}
\date{\today} \title{Chiral Spin Liquids and Quantum Error Correcting
Codes} \author{N.~E.~Bonesteel} \address{National High Magnetic Field
Laboratory and Department of Physics, Florida State University,
Tallahassee, FL 32310} \maketitle
\begin{abstract}
The possibility of using the two-fold topological degeneracy of
spin-1/2 chiral spin liquid states on the torus to construct quantum
error correcting codes is investigated. It is shown that codes
constructed using these states on finite periodic lattices do not meet
the necessary and sufficient conditions for correcting even a single
qubit error with perfect fidelity. However, for large enough lattice
sizes these conditions are approximately satisfied, and the resulting
codes may therefore be viewed as approximate quantum error correcting
codes.

\end{abstract}


\begin{multicols}{2}

\section{Introduction}

If one could be built, a quantum computer would be capable of solving
certain computational problems much more efficiently than any
classical computer, most notably factoring large integers into primes
in polynomial time \cite{shor1} and searching unordered lists of $N$
items in $O(N^{1/2})$ queries \cite{grover}. In a quantum computer
classical bits, which take the values 0 or 1, are replaced by quantum
bits, or qubits --- two-level quantum systems whose Hilbert spaces are
spanned by the orthonormal states $|0\rangle$ and $|1\rangle$.  Unlike
a classical bit, a qubit can therefore be placed in an arbitrary
quantum superposition $\alpha |0\rangle + \beta |1\rangle$.  However,
due to the coupling of this qubit with the outside world, which may be
small but which can never be reduced to zero, this state will
eventually become entangled with its environment, losing its quantum
coherence.  Because maintaining this coherence is crucial for quantum
computers to achieve their superiority over classical computers, the
question of how to protect qubits from decoherence has been central to
the ongoing effort in quantum computing.

One of the most surprising recent developments in quantum information
theory has been the discovery of a scheme for fighting decoherence
using what are called quantum error correcting codes
\cite{shor2,steane}. A quantum error correcting code is a mapping from
the Hilbert space of a single qubit, (or, possibly, more than one
qubit), to a subspace of the Hilbert space of many physical qubits.
The resulting many qubit state is then referred to as an {\it encoded}
qubit.  These encoded states are carefully designed so that if an
error occurs, i.e., if a small number of the physical qubits become
entangled with their environment, certain measurements can be
performed to determine which error has occurred and how it can be
corrected {\it without} disturbing the quantum information stored in
the encoded qubit.

An important connection between quantum error correcting codes,
many-body physics and topological quantum numbers was pointed out by
Kitaev \cite{kitaev1} who constructed a class of spin Hamiltonians
realized on two-dimensional lattices with periodic boundary
conditions.  For the simplest of these models the physical qubits
correspond to spin-1/2 particles located on lattice edges.  Kitaev
showed that these models have degenerate ground states which are
distinct, orthogonal states but which nevertheless cannot be
distinguished by any {\it local} operators.  Instead, these degenerate
states can only be distinguished by {\it global} operators, e.g.,
operators which act on a set of spins which encircle a topologically
nontrivial orbit of the torus formed by the periodic lattice.  Kitaev
showed that these `topologically' degenerate states could be used as
quantum error correcting codes, called toric codes, in which encoded
qubits correspond to particular superpositions of these degenerate
ground states.  Because these states can only be distinguished by
topological quantum numbers, and not by any local observables, the
environment has difficulty `measuring' the encoded quantum information
which is therefore, to some extent, protected from decoherence.

This connection between quantum spin models with topologically
degenerate ground states and quantum error correcting codes provides
motivation to revisit some of the topological degeneracies which can
arise `naturally' in certain condensed matter systems.  In this paper,
as a concrete example, the spin-1/2 chiral spin liquid states
originally proposed by Kalmeyer and Laughlin \cite{kalmeyer} as
possible spin liquid ground states for frustrated spin-1/2 quantum
antiferromagnets are considered in this context.  It is known that
these states possess a kind of topological order \cite{wen2}, not
unlike the topological order of Kitaev's toric codes, which leads to
nontrivial ground state degeneracies on Riemann surfaces with genus 1
or greater.  It should be emphasized that it is by no means clear that
the results of this paper will be useful for constructing quantum
error correction schemes for realistic quantum computers. Rather, the
goal of the present work is to provide some insight into the possible
ways that the Hilbert space of an array of qubits can exhibit
topological quantum numbers.

The paper is organized as follows.  In Section II the basic physics of
the chiral spin liquid states is reviewed.  The case of finite $N_1
\times N_2$ periodic lattices is considered and it is proven that, if
properly constructed, the chiral spin liquid states realized on these
lattices are exact singlet states, generalizing a previous proof due
to Laughlin that these states are singlets for $N \times N$ periodic
lattices with $N$ even \cite{laughlin1}. In addition, it is shown by
explicit construction that these states possess a topological
degeneracy on any periodic lattice, in agreement with \cite{wen2}.  In
Section III the nature of this topological degeneracy is characterized
using the Lieb-Schultz-Mattis `slow twist' operator and it is shown to
be related to a topological decoupling of the Hilbert space of
short-range valence-bond states on periodic lattices.  In Section IV
the general properties of quantum error correcting codes are reviewed
and compared with analogous properties of the topologically degenerate
chiral spin liquid states, which are computed by variational Monte
Carlo.  It is shown that, unlike the toric codes, chiral spin liquid
states do {\it not} satisfy the necessary and sufficient conditions to
be exact quantum error correcting codes.  However, as the lattice size
increases, the violation of these conditions becomes weaker until, in
the thermodynamic limit, they become satisfied.  In this sense the
topologically degenerate chiral spin liquid states may be viewed as
approximate quantum error correcting codes on large enough lattices.
Finally, Section V summarizes the results and conclusions of the
paper.

\section{Chiral Spin Liquid Wave Functions on Finite Periodic Lattices}

Consider the following spin-1/2 Hamiltonian realized on a
two-dimensional square lattice,
\begin{eqnarray}
H = J_1 \sum_{\langle i,j \rangle} {\bf S}_i \cdot {\bf S}_j + J_2
\sum_{\langle\langle i,j \rangle\rangle} {\bf S}_i \cdot {\bf S}_j ,
\label{spinhamiltonian}
\end{eqnarray}
where $J_1, J_2 \ge 0$ and $\langle i,j \rangle$ and $\langle \langle
i,j \rangle \rangle$ denote nearest-neighbor and next-nearest-neighbor
pairs of lattice sites, respectively.  The lattice size is taken to be
$N_1 \times N_2$ with $N_1$ even and lattice spacing $b$. Periodic
boundary conditions will be assumed throughout the paper.  For
concreteness (and future reference) the lattice is taken to lie in the
$xy$ plane, with lattice sites ${\bf r} = (n_1 {\hat {\bf x}} + n_2
{\hat {\bf y}})b$ where $n_1$ and $n_2$ are integers.

If $J_1 > 0$ and $J_2 = 0$ Hamiltonian (\ref{spinhamiltonian})
describes an unfrustrated two-dimensional spin-1/2 Heisenberg
antiferromagnet for which the ground state is known to possess
long-range N\'eel order in the thermodynamic limit.  In the opposite
extreme $J_2 \gg J_1$ the two sublattices decouple, and each develops
N\'eel order independently.  It is generally believed that over an
intermediate range of $J_2/J_1$ values the ground state is in a
`spin-peierls' phase with a locally observable broken translational
symmetry, but there is no evidence that Hamiltonian
(\ref{spinhamiltonian}) ever has a spin liquid ground state, i.e., a
ground state with neither long-range N\'eel order nor any other
locally observable broken translational symmetry, other than at zero
temperature critical points.  Nevertheless, in what follows
Hamiltonian (\ref{spinhamiltonian}) will be used to introduce the
chiral spin liquid states with the understanding that while these
states almost certainly do not describe the ground state of
(\ref{spinhamiltonian}) in any parameter range, they {\it may} be
eigenstates of an, as yet unknown, frustrated spin Hamiltonian.
Fortunately, the topic of this paper -- the relationship between
chiral spin liquid states and quantum error correcting codes --
involves properties of Hilbert space and does not depend on the
Hamiltonian.

Hamiltonian (\ref{spinhamiltonian}) can be viewed as describing a
system of $N$ interacting hard core bosons hopping on a square lattice
where $N = N_1 N_2/2$. In this description the bosons correspond to up
spins moving in a down spin background with matrix elements $J_1$ and
$J_2$ for nearest-neighbor and next-nearest-neighbor hopping,
respectively. If the totally symmetric wave function describing these
bosons is $\Phi(\{{\bf r}_i\})$ then the corresponding spin state is
\begin{eqnarray}
|\Phi\rangle = \sum_{\{ {\bf r}_1 {\bf r}_2 \cdots {\bf r}_N \}}
\Phi(\{{\bf r}_i\}) S^+_{{\bf r}_1} S^+_{{\bf r}_2} \cdots S^+_{{\bf
r}_N} |\downarrow\downarrow\cdots\downarrow\rangle.
\end{eqnarray}

Because $J_1$ and $J_2$ are both positive the effective hopping for
the bosons is frustrated.  As pointed out by Kalmeyer and Laughlin
\cite{kalmeyer} this frustration can be viewed as being due to the
presence of a magnetic field.  To see this imagine that each bosons
has charge $q$ and moves in the presence of a magnetic field
perpendicular to the plane of the lattice with field strength $B =
\phi_0/b^2$ where $\phi_0 = hc/q$ is the flux quantum.  The
corresponding vector potential in the the Landau gauge is then ${\bf
A} = -By {\hat {\bf x}}$ and it can readily be shown that
\begin{eqnarray} J_1 &=& J_1 \exp
\frac{i2\pi}{\phi_0} \int_{{\bf r}_i}^{{\bf r}_j} {\bf A}\cdot d{\bf
l},\\ J_2 &=& -J_2 \exp \frac{i2\pi}{\phi_0} \int_{{\bf r}_i}^{{\bf
r}_j} {\bf A}\cdot d{\bf l},
\end{eqnarray}
where ${\bf r}_i$ and ${\bf r}_j$ denote the starting and ending sites
of the relevant hopping process and the line integrals are taken along
straight lines connecting these sites.  Thus a positive (frustrated)
$J_2$ corresponds to a negative (unfrustrated) $J_2$ in the presence
of a fictitious magnetic field of suitable strength.

The sign of $J_1$ can be changed without affecting $J_2$ by dividing
the square lattice into $A$ and $B$ sublattices and rotating the spins
on the $A$ sublattice by $2\pi$ radians about any fixed axis in spin
space while leaving the $B$ sublattice untouched.  Under this
sublattice rotation the boson wave function is transformed according
to
\begin{eqnarray}
\Psi(\{{\bf r}_i\}) = \left(\prod_i e^{i b (x_i+y_i)/2} \right)
\Phi(\{{\bf r}_i\}).
\label{sublattice}
\end{eqnarray}
It is important for what follows to note that the spin wave function
$\Phi$ must satisfy periodic boundary conditions on an $N_1 \times
N_2$ lattice in the $x$ and $y$ directions where $N_1$ is even.
Therefore for even values of $N_2$ the transformed wave function
$\Psi$ must also satisfy periodic boundary conditions in the $x$ and
$y$ directions, while for odd values of $N_2$, due to the sublattice
mismatch, $\Psi$ must satisfy periodic boundary conditions in the $x$
directions and {\it antiperiodic} boundary conditions in the $y$
direction.

This mapping from a frustrated spin model to hard core bosons hopping
on a lattice in the presence of a magnetic field inspired Kalmeyer and
Laughlin to propose a trial wave function based on the related system
of interacting bosons moving in free space \cite{kalmeyer}.  Following
their work, imagine that no lattice is present and that the bosons
move on a torus of length $L_1 = N_1 b$ in the $x$ direction and $L_2
= N_2 b$ in the $y$ direction.  Since the magnetic field is $B =
\phi_0/b^2 = 2 (hc/q) n$, where $n$ is the number density of bosons,
the effective Landau level filling fraction for the bosons is
$\nu=1/2$.  In what follows the effective magnetic length $l_0 =
(\hbar c/qB)^{1/2}$ is taken as the natural length scale and set equal
to 1, so that, for example, the lattice spacing is $b=(2\pi)^{1/2}$.

A natural {\it Ansatz} for the ground state of this many-boson system
with strong short-range repulsion is that the bosons condense into a
$\nu=1/2$ bosonic Laughlin state.  The corresponding Laughlin wave
function is completely determined by the lowest Landau level
constraint and the requirement that the wave function vanish as $\sim
(z_i - z_j)^2$ as two bosons approach one another. In the Landau gauge
the Laughlin wave function for $\nu=1/2$ bosons on an $L_1 \times L_2$
torus can be written \cite{haldane2}
\begin{eqnarray}
\Psi(\{{\bf r}_i\}) = \psi(\{z_i\}) \prod_ie^{-y_i^2/2},
\end{eqnarray}
where
\begin{eqnarray}
\psi(\{ z_l\}) = F(Z) \prod_{i<j} \vartheta_1(\pi(z_i -
z_j)/L_1 | \tau)^2.
\end{eqnarray}
Here $\vartheta_1(z|\tau)$ is the odd elliptic theta function
\cite{gradshteyn} with $\tau = iL_2/L_1$, $z_i = x_i + i y_i$ is the
complex coordinate of the $i$th boson, $Z = \sum_i z_i$ is the center
of mass coordinate, and 
\begin{eqnarray}
F(Z) = e^{iKZ}\prod_{\nu=1}^2 \vartheta_1(\pi(Z - w_\nu)/L_1 |\tau)
\label{fofz}
\end{eqnarray}
is the center of mass part of the wave function.

The constants $K$, $w_1$ and $w_2$ in (\ref{fofz}) must be chosen so
that $\Psi$ satisfies the twisted boundary conditions
\begin{eqnarray}
\Psi({\bf r}_1+{\hat{\bf x}} L_1,{\bf r}_2,\cdots) &=&
e^{i\phi_1}\Psi({\bf r}_1,{\bf r}_2,\cdots) , \\ \Psi({\bf
r}_1+{\hat{\bf y}} L_2,{\bf r}_2,\cdots) &=& e^{i\phi_2}e^{-iL_2
x}\Psi({\bf r}_1,{\bf r}_2,\cdots) ,
\label{bc2}
\end{eqnarray}
for each boson coordinate.  Here $\phi_1$ and $\phi_2$ are two
toroidal fluxes which characterize the $x$ and $y$ boundary
conditions.  The requirement that the wave function satisfy these
boundary conditions leads to the following restrictions on $K$, $w_1$
and $w_2$ \cite{haldane2},
\begin{eqnarray}
e^{iKZ} &=& e^{i\phi_1} , \label{rest1}
\\ e^{i2\pi (w_1+w_2)/L_1} &=& e^{i\phi_2} e^{kL_2} ,
\label{rest2}
\end{eqnarray}
which can be satisfied in a variety of ways \cite{haldane2}.  Here the
$K=0$ `coherent states' are used, for which
\begin{eqnarray}
F_n(Z) = \vartheta_1 (\pi(Z - W_n)/L_1|\tau)^2 ,
\end{eqnarray}
where
\begin{eqnarray}
W_n = \left(\frac{n}{2} + \frac{\phi_2}{4\pi}\right)L_1 ,
\end{eqnarray}
for $n=0$ and 1.  

The two degenerate Laughlin states, $\Psi_0$ and $\Psi_1$,
corresponding to $n=0$ and 1, are distinguished only by the difference
in the center of mass parts of their wave functions, $F_n(Z)$.  As
shown by Haldane \cite{haldane1} this two-fold degeneracy is required
for any translationally invariant system on a torus at $\nu=1/2$. Note
that although the states $\Psi_0$ and $\Psi_1$ span the
two-dimensional Hilbert space of Laughlin states on the torus, they
are not orthogonal.  However, when $L_1 \gg L_2$ it can be shown that
\begin{eqnarray}
F_n(Z) \simeq \sqrt{\frac{L_2}{L_1}} \sum_{m=-\infty}^\infty \exp
-\frac{\pi (Z-W_n - L_1(m+\frac{1}{2}))^2}{L_1 L_2}.
\end{eqnarray}
This function is sharply peaked at the points $Z = W_n + L_1/2 + mL_1$
for any integer $m$ with peak widths $\sim \sqrt{L_1 L_2}$.  It
follows that in the limit $L_1/L_2 \rightarrow \infty$ the overlap
between $F_0$ and $F_1$ vanishes and the states $\Psi_0$ and $\Psi_1$
become orthogonal.

To use $\Psi_0$ and $\Psi_1$ as wave functions for hard core bosons on
the $N_1 \times N_2$ periodic square lattice the boson coordinates are
restricted to the lattice points ${\bf r} = (n_1 {\hat {\bf x}} + n_2
{\hat {\bf y}})b$.  For these lattice points the $e^{-iL_2x}$ factor
in (\ref{bc2}) is identically 1 and the toroidal fluxes $\phi_1$ and
$\phi_2$ then correspond to overall phases associated with the
boundary conditions in the $x$ and $y$ directions.  As shown above,
for even and odd values of $N_2$ the boson wave functions $\Psi_n$ are
required to be, respectively, periodic and antiperiodic, in the $y$
direction.  Thus $\phi_2 = 0$ for even values of $N_2$ and $\phi_2 =
\pi$ for odd values of $N_2$, implying that the parameter $W_n$
appearing in $F_n(Z)$ is
\begin{eqnarray}
W_n = \left\{
\begin{array}{cc}
n L_1/2 & N_2\ {\rm even}, \\ (2n+1)L_1/4 & N_2\
{\rm odd}.
\end{array}
\right.
\label{wn}
\end{eqnarray}
Because $N_1$ is assumed to be even the wave functions $\Psi_n$ are
always required to be periodic in the $x$ direction implying that
$\phi_1 = 0$ for all lattices.

The spin states $\Phi_n$ obtained from $\Psi_n$ by undoing the
sublattice rotation (\ref{sublattice}) are referred to as chiral spin
liquid states.  These states break both time-reversal symmetry (T) and
parity (P) under both of which $\Phi \rightarrow \Phi^*$ \cite{wen1}.
This broken symmetry, characterized by a nonvanishing chiral order
parameter $\langle {\bf \sigma}_1 \cdot ({\bf \sigma}_2 \times {\bf
\sigma}_3)\rangle$, leads to a two-fold degeneracy which is clearly
{\it not} topological because it can be identified by measuring a
local order parameter, and therefore not potentially useful for
constructing quantum error correcting codes. However, the additional
degeneracy associated with the center of mass part of the wave
function is global in character and not associated with any local
order parameter (at least in the thermodynamic limit, see Section IV).
Using a field theoretic description, Wen has argued that the
degeneracy of chiral spin liquids on a two-dimensional closed surface
with genus $g$ should be $2 (2)^g$ where the overall factor of 2 is
due to the broken $T$ and $P$ symmetries and the factor of $2^g$ is a
measure the topological order \cite{wen2}.  The explicit construction
of the states $\Phi_0$ and $\Phi_1$ on periodic lattices is therefore
consistent with Wen's result for $g=1$.

The requirement that the spin states $\Phi_n$ be singlet states is
equivalent to the requirement $S^- |\Phi_n\rangle = 0$ which in turn
implies that
\begin{eqnarray}
{\sum_{\bf r_1}}^\prime \Phi_n({\bf r}_1,{\bf r}_2,\cdots) = 0 ,
\label{singlet1}
\end{eqnarray}
where the primed sum denotes a sum over lattice sites on the torus.
If the toroidal fluxes are chosen so that $\Phi_n$ is periodic in both
the $x$ and $y$ directions when evaluated on lattice sites then it can
be shown that $\Phi_n$ satisfies (\ref{singlet1}) by using the singlet
sum rule derived by Laughlin \cite{laughlin1},
\begin{eqnarray}
\sum_{{\bf r}} G({\bf r}) f(z) e^{-|z|^2/2} = 0 ,
\label{sumrule}
\end{eqnarray}
where for lattice sites ${\bf r} = (n_1\hat{\bf x} + n_2\hat{\bf y}) b$,
\begin{eqnarray}
G({\bf r}) = (-1)^{n_1n_2 + n_1 + n_2 + 1},
\end{eqnarray}
and $f(z)$ is any polynomial in $z$.  Note that in order for
(\ref{sumrule}) to be satisfied it is necessary to sum over all
lattice points on the infinite two-dimensional plane.

Following Laughlin \cite{laughlin1} the sum rule (\ref{sumrule}) can
be applied to the chiral spin liquid wave functions for finite $N_1
\times N_2$ periodic lattices by first exploiting the periodicity of
$\Phi$ in the $x$ and $y$ directions to extend the summation in
(\ref{singlet1}) to the entire lattice,
\begin{eqnarray}
{\sum_{{\bf r}_1}}^\prime \Phi_n({\bf r}_1,{\bf r}_2,\cdots) &=&
\lim_{R\rightarrow \infty} \frac{2\pi N_1 N_2}{\pi R^2} \sum_{|{\bf r}_1| < R}
\Phi_n({\bf r}_1,{\bf r}_2,\cdots),
\end{eqnarray}
and then using the following identity which holds for all lattice
points,
\begin{eqnarray}
e^{ib(x+y)/2} e^{-y^2/2} = - G({\bf r}) e^{z^2/4}e^{-|z|^2/4} ,
\end{eqnarray}
to show that
\begin{eqnarray}
&&\sum_{|{\bf r}_1| < R} \Phi_n({\bf r}_1,{\bf r}_2,\cdots)\nonumber\\
&& ~ = \sum_{|{\bf r}_1| < R} G({\bf r}_1) e^{z_1^2/4} F_n(Z)
\psi(z_1,z_2,\cdots) e^{-|z_1|^2/4}\prod_{i\ne 1}e^{-y_i^2/2}.\nonumber\\
\label{sing}
\end{eqnarray}
In the limit $R \rightarrow \infty$ the summation on the right hand
side of (\ref{sing}) vanishes due to the sum rule (\ref{sumrule}) and
the fact that the function $e^{z_1^2/4} F_n(Z) \psi(z_1,z_2,\cdots)$
is analytic in $z_1$.  The chiral spin liquid states are therefore
singlets for {\it any} $N_1 \times N_2$ lattice where $N_1$ is even,
provided the toroidal fluxes have been chosen, as they have been here,
to ensure that the spin wave function is periodic in both the $x$ and
$y$ directions.

The nature of the topological degeneracy of the chiral spin liquid
states depends on whether $N_2$ is even or odd.  To understand this
distinction consider the translation operator $T_x$ which translates
each boson by one lattice vector in the $x$ direction.  Under this
operator the relative part of the wave function $\Phi_n$ is unaffected
and only the center of mass coordinate is shifted according to
\begin{eqnarray}
Z \rightarrow Z+N b = Z+ \frac{N_1 N_2 b}{2} = Z+\frac{N_2 L_1}{2}.
\end{eqnarray}
For odd values of $N_2$ this implies that the center of mass is
shifted through a half-odd integer multiple of $L_1$ and, due to the
periodic boundary conditions, this is equivalent to a net shift of the
center of mass by $L_1/2$.  Thus, in obvious notation,
\begin{eqnarray}
T_x F_n(Z) = F_n(Z+L_1/2),
\end{eqnarray}
from which it follows that
\begin{eqnarray}
T_x \Phi_0 = \Phi_1~~~~~{\rm and}~~~~~~T_x \Phi_1 = \Phi_0.
\end{eqnarray}
This implies that for odd values of $N_2$ the spin liquid states
$\Phi_0$ and $\Phi_1$ break translational symmetry.  In contrast, for
even values of $N_2$, the center of mass coordinate is shifted through
an integer multiple of $L_1$ and the translation operator $T_x$ has no
effect,
\begin{eqnarray}
T_x \Phi_0 = \Phi_0~~~~~{\rm and}~~~~~~T_x \Phi_1 = \Phi_1.
\end{eqnarray}
Finally, because $N_1$ is even, for both even and odd values of $N_2$
\begin{eqnarray}
T_y \Phi_0 = \Phi_0~~~~~{\rm and}~~~~~~T_y \Phi_1 = \Phi_1,
\end{eqnarray}
where $T_y$ is the translation operator which translates each boson by
one lattice vector in the $y$ direction.

\section{
Connection to Lieb-Schultz-Mattis Operator and Valence-Bond Topology}

The topological degeneracy of the chiral spin liquid states can be
elucidated further by introducing Affleck's two-dimensional
generalization of the Lieb-Schultz-Mattis slow twist operator
\cite{lieb,affleck},
\begin{eqnarray}
U_{LSM} = \exp \left(i \frac{\pi}{L_1} {\sum_{\bf r}}^\prime x
\sigma_{\bf r}^z\right),
\end{eqnarray}
where the primed sum denotes a sum over lattice points on the torus.
The usefulness of this operator derives partly from the fact that when
$N_1 \gg N_2$, for any singlet state $|{\rm Sing}\rangle$ and any
rotationally invariant spin Hamiltonian such as
(\ref{spinhamiltonian}) which only includes short-range interactions,
it can be shown that \cite{lieb,affleck}
\begin{eqnarray}
\langle {\rm Sing}| \left( U_{LSM} H U_{LSM}^\dagger - H \right)| {\rm
Sing} \rangle \sim O\left(J\frac{N_2}{N_1}\right),
\label{lsmprop}
\end{eqnarray}
where $J$ is a measure of the typical magnetic interaction strength.
If, as is supposed to be the case here, $H$ has degenerate singlet
ground states separated by a gap from all excited states, then
(\ref{lsmprop}) implies that in the $N_1/N_2 \rightarrow\infty$ limit
$U_{LSM}$ maps states in the finite dimensional Hilbert space spanned
by these states into one another.

The Lieb-Schultz-Mattis slow twist operator can be recast in bosonic
language as
\begin{eqnarray} U_{LSM}  = \exp \left(-i
\frac{\pi}{L_1} {\sum_{\bf r}}^\prime x \right) \exp \left(i
\frac{2\pi}{L_1} X \right),
\label{uboson}
\end{eqnarray}
where $X$ is the $x$ coordinate of the center of mass.  Due to the
periodic boundary conditions there is some freedom in labeling the
lattice sites on the torus, and in order to precisely define $U_{LSM}$
it is necessary to choose a particular labeling scheme. Here it will
be assumed that the primed sum in (\ref{uboson}) is over lattice sites
${\bf r} = (n_1 {\hat {\bf x}} + n_2 {\hat {\bf y}})b$ where $n_1 =
-N_1/2+1,\cdots,N_1/2$ and $n_2 = 1,\cdots, N_2$.  For this choice
$\sum_{\bf r}^\prime x = L_2 L_1/(2b)$ and
\begin{eqnarray}
U_{LSM} = (-i)^{N_2} \exp \left(i \frac{2\pi}{L_1} X \right).
\label{iulsm}
\end{eqnarray}

As shown in Section II, when $L_1 \gg L_2$ the center of mass part of
the chiral spin liquid wave functions $\Phi_n$ becomes sharply peaked
for $Z = W_n + L_1/2 + m L_1$ for any integer $m$.  Therefore in this
limit
\begin{eqnarray}
\exp \left( i\frac{2\pi}{L_1} X \right) F_n(Z) \simeq - \exp \left(
i\frac{2\pi}{L_1}W_n \right) F_n(Z).
\label{comfac}
\end{eqnarray}
Combining (\ref{iulsm}) and (\ref{comfac}) and using (\ref{wn}) one
finds that in the $N_1/N_2 \rightarrow \infty$ limit the states
$\Phi_0$ and $\Phi_1$ become eigenstates of $U_{LSM}$ with eigenvalues
$\pm 1$,
\begin{eqnarray}
\lim_{\frac{N_1}{N_2} \rightarrow \infty} U_{LSM} \Phi_n = \left\{
\begin{array}{lc}
(-1)^n (-1)^{\frac{N_2+2}{2}} \Phi_n & N_2\ {\rm even},\\ \\
(-1)^n (-1)^{\frac{N_2+1}{2}} \Phi_n & N_2\ {\rm odd}.
\end{array}
\right.
\label{ueigenvalue}
\end{eqnarray}

\begin{figure}[htb]
\centerline{\psfig{figure=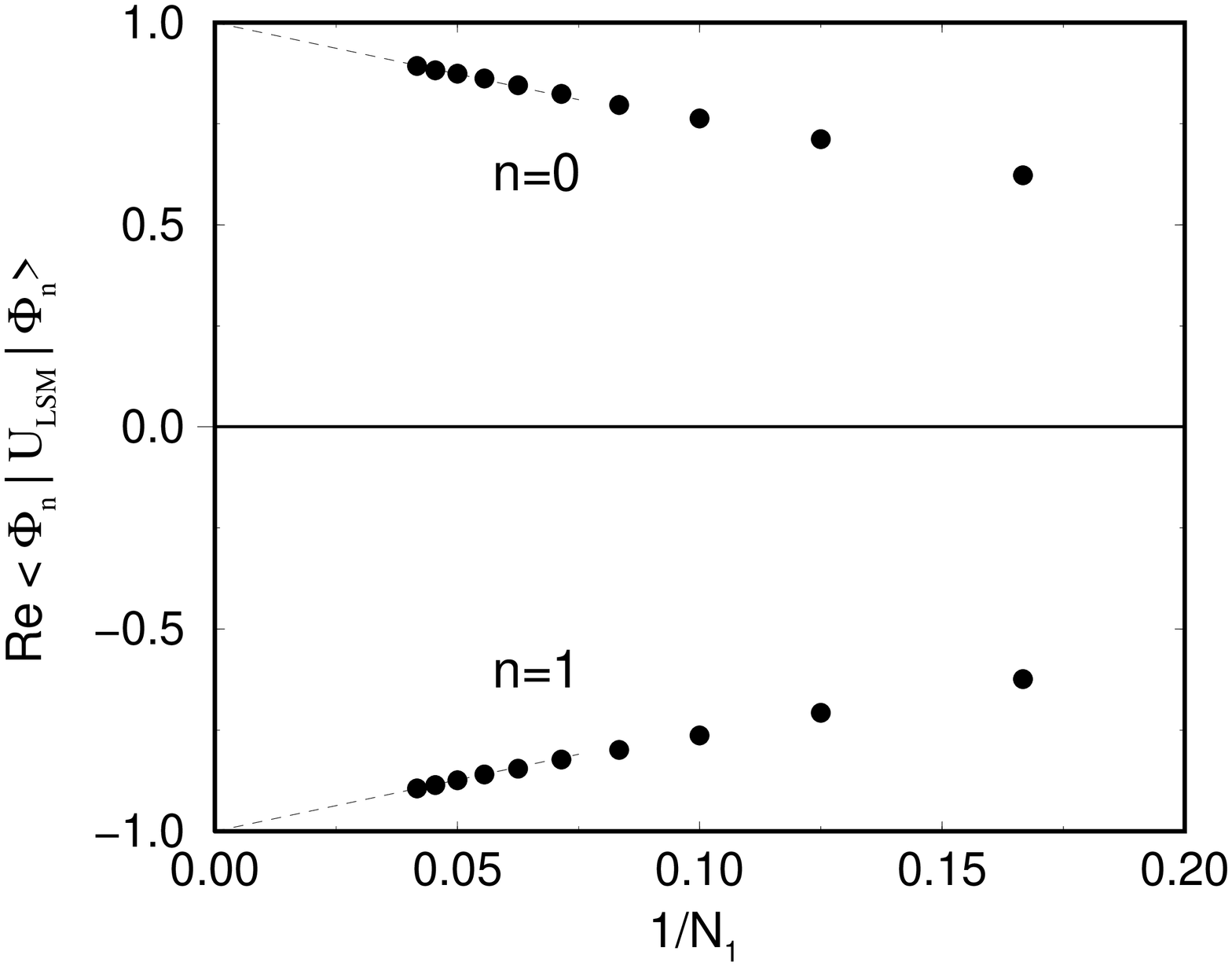,height=2.5in,angle=270}}
\centerline{\psfig{figure=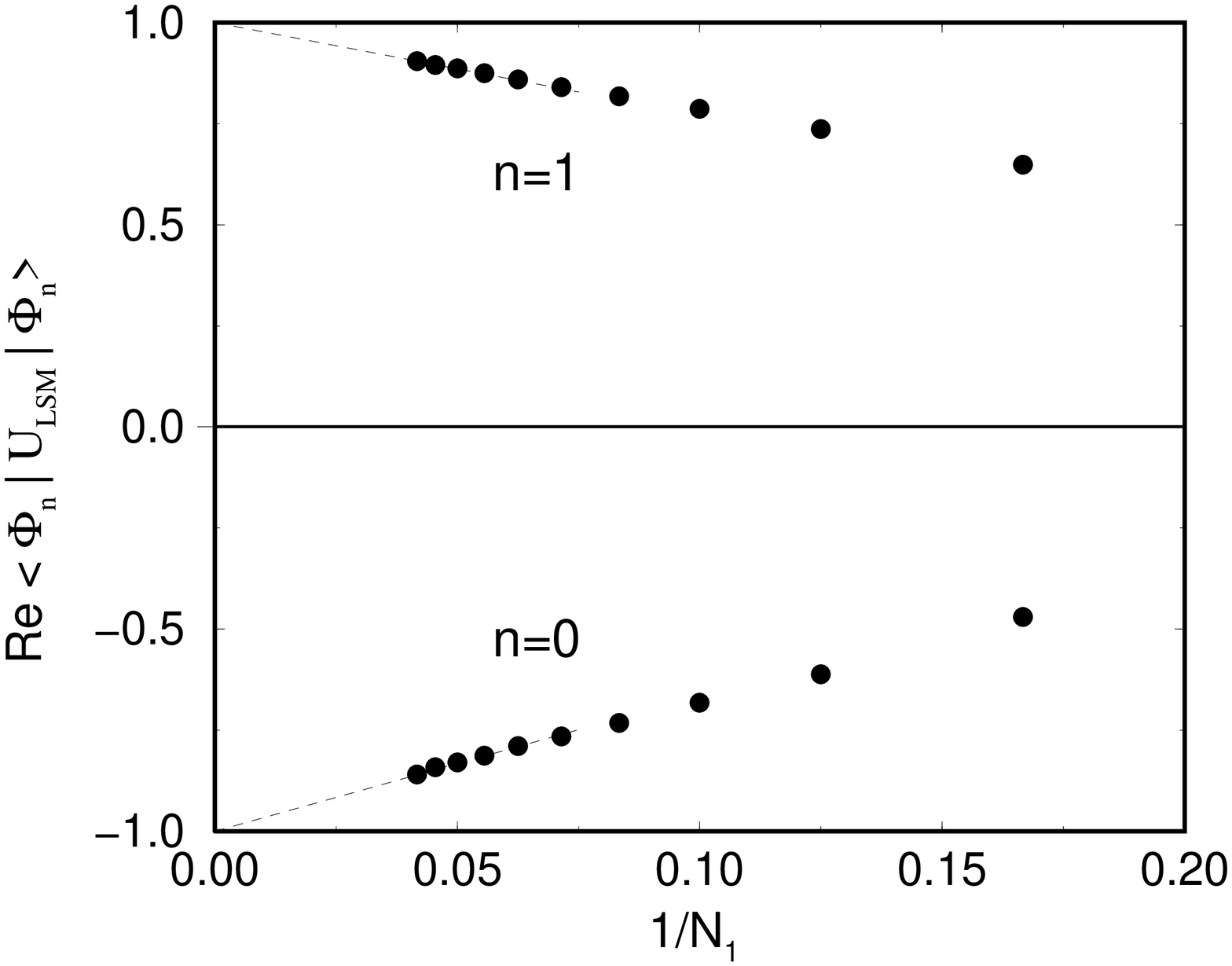,height=2.5in,angle=270}}
\vskip .15in
\caption{Real part of the expectation value of the Lieb-Schultz-Mattis
slow twist operator in the chiral spin liquid states $\Phi_0$ and
$\Phi_1$ for $N_1 \times 3$ lattices (top) and $N_1 \times 4$ lattices
(bottom) plotted vs $1/N_1$.  Statistical error bars are smaller than
symbol sizes.}
\label{uexp}
\end{figure}

Figure \ref{uexp} shows the results of a numerical variational Monte
Carlo computation of the real part of the expectation values $\langle
\Phi_n | U_{LSM} | \Phi_n \rangle$ for $n=0$ and 1 and $N_2 = 3$ and 4
plotted vs $1/N_1$.  The results clearly show that ${\rm Re}\langle
\Phi_n | U_{LSM} | \Phi_n \rangle \rightarrow \pm 1$ in the $N_1
\rightarrow \infty$ limit for fixed $N_2$.  Because $U_{LSM}$ is a
unitary operator it follows that $\Phi_0$ and $\Phi_1$ become
eigenstates of $U_{LSM}$ in the $N_1/N_2\rightarrow\infty$ limit with
eigenvalues $\pm 1$, consistent with (\ref{ueigenvalue}).

The fact that $\Phi_0$ and $\Phi_1$ become eigenstates of $U_{LSM}$ in
the $N_1/ N_2 \rightarrow \infty$ limit leads to an appealing picture
of the topological degeneracy of chiral spin liquids in terms of the
valence-bond state basis.  This basis consists of states in which
spins are singlet correlated in pairs, which are said to be connected
by valence bonds.  While {\it any} singlet state may be represented as
a linear superposition of valence-bond states, it is reasonable to
assume that any singlet state, such as the chiral spin liquid states,
in which the spin-spin correlation function decays exponentially with
distance \cite{kalmeyer,laughlin1} can by represented as a
superposition of {\it short-range} valence-bond states.  A short-range
valence-bond state is a valence-bond state containing only bonds with
lengths less than a specified length, or bonds with a distribution of
lengths which falls off exponentially for long bonds.

\begin{figure}[htp]
\centerline{\psfig{figure=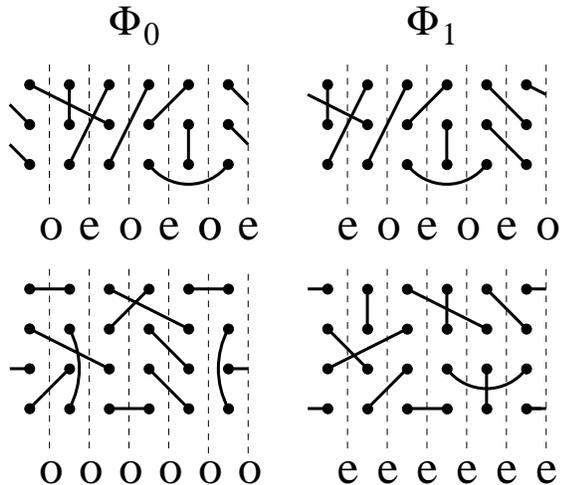,height=2.5in,angle=270}}
\vskip .15in
\caption{Four short-range valence-bond configurations illustrating the
topological quantum numbers responsible for the degeneracy of chiral
spin liquid states on periodic lattices. Dashed lines are drawn
through horizontal gaps in these configurations with the parity of the
number of valence bonds crossed by each line shown below (e=even, o =
odd). The upper two configurations on $6 \times 3$ lattices indicate
the generic behavior for odd width lattices in which an alternating
even-odd or odd-even pattern appears. The lower two configurations on
$6 \times 4$ lattices indicate the generic behavior for even width
lattices in which the gap parities are either all odd or all even.
The configurations on the left and right contribute, respectively, to
the states $\Phi_0$ and $\Phi_1$, in a sense described in the text.}
\label{lsm}
\end{figure}

The requirement that valence bonds must connect two sites, and only
one bond may be attached to each site, gives rise to a topological
decoupling of the space of short-range valence-bond states
\cite{thouless,rokhsar,read,bonesteel}.  Figure \ref{lsm} shows four
short-range valence-bond states, two on a $6 \times 3$ lattice and two
on a $6 \times 4$ lattice. In this figure solid lines connecting pairs
of lattice sites represent valence bonds. In each of these states the
$x$ projection of the length of each bond does not exceed $2 b$ and so
it is possible to unambiguously determine the way in which a given
bond `wraps' around the periodic boundary condition in the $x$
direction (it is in this sense that these states are short-range
valence-bond states).  For each of these states 6 vertical dashed
lines are shown which `slice' the gaps between each vertical line of
lattice sites.  The parity (o = odd, e = even) of the number of bonds
cut by these dashed lines is then shown below each line.

For $N_2 = 3$, or any odd value of $N_2$, an alternating even-odd
pattern invariably appears \cite{bonesteel}.  Short-range valence-bond
states then fall into two distinct classes, which can be referred to
as even-odd and odd-even, corresponding to the two $N_2=3$
configurations shown in Fig.~\ref{lsm}.  For $N_2 = 4$, or any even
value of $N_2$, all gaps have the same parity and, again, there are
two possibilities, either each gap has odd parity or each gap has even
parity, corresponding to the two $N_2 = 4$ states shown in
Fig.~\ref{lsm}.  It is convenient to define a topological quantum
number, the gap parity, of a given short-range valence-bond state
$|\alpha\rangle$ to be $(-1)^{\gamma_\alpha}$ where $\gamma_\alpha$ is
the number of bonds which cross the gap between the line of lattice
points with $x=L_1/2$ and those with $x=1-L_1/2$, i.e., those bonds
which cross the discontinuity in $x$ due to the periodic boundary
conditions using the site labeling scheme introduced above.  According
to this definition, for the two configurations at the top of
Fig.~\ref{lsm} $\gamma_\alpha = 2$ and 1, and the gap parities are
$+1$ and $-1$, while for the two configurations at the bottom of
Fig.~\ref{lsm} $\gamma_\alpha = 1$ and 2, and the gap parities are
$-1$ and $+1$.

In \cite{bonesteel} it was shown that if $N_1 \gg N_2$ for any
short-range valence-bond state $|\alpha\rangle$
\begin{eqnarray}
U_{LSM} | \alpha \rangle \simeq (-1)^{\gamma_\alpha} | \alpha \rangle.
\label{vbalsmt}
\end{eqnarray}
The appearance of the gap parity, $(-1)^{\gamma_\alpha}$, in
(\ref{vbalsmt}) is due to the minus sign obtained whenever a spin-1/2
particle is rotated through $2\pi$ radians about any axis.  If a given
short-range valence-bond state is acted on by $U_{LSM}$ then, if $N_1
\gg N_2$, for most valence bonds in $|\alpha\rangle$ the two spins
forming the bond are rotated by approximately the same amount.  These
valence bonds are therefore only weakly affected by the slow twist
operator.  However, for those $\gamma_\alpha$ bonds which cross the
discontinuity in $x$ due to the periodic boundary conditions the
operator $U_{LSM}$ rotates one spin by approximately $2\pi$ radians
while the other spin is, again approximately, not rotated at
all. Therefore, while these bonds also remain approximately in singlet
states, they each contribute a factor of $-1$ to (\ref{vbalsmt})
because only {\it one} spin has been rotated through $2\pi$ radians.

According to (\ref{ueigenvalue}), when $N_1 \gg N_2$ the states
$\Phi_0$ and $\Phi_1$ become eigenstates of $U_{LSM}$ with eigenvalues
$\pm 1$. It is therefore plausible to assume that these states can be
represented as linear superpositions of those short-range valence-bond
states which also become eigenstates of $U_{LSM}$ in this limit with
the same eigenvalues, i.e., those states whose gap parities are equal
to the corresponding eigenvalues given in (\ref{ueigenvalue}).  The
gap parity can then be viewed as the topological quantum number which
distinguishes between the states $\Phi_0$ and $\Phi_1$ in this limit.
Note that the alternating even-odd or odd-even patterns which appear
in the gap parities for odd values of $N_2$, and the uniform gap
parities, either all even or all odd, which appear for even values of
$N_2$, are consistent with the symmetry properties of $\Phi_0$ and
$\Phi_1$ under the translation operators $T_x$ derived in Sec.~II.
For more details on the connection between the Lieb-Schultz-Mattis
slow twist operator and the topological decoupling of short-range
valence-bond states see \cite{bonesteel}.

\section{Connection to Quantum Error Correcting Codes}

In the previous section it was shown that, in a sense which becomes
precise in the limit $N_1 \gg N_2$, the topological quantum number
distinguishing degenerate chiral spin liquid states is the gap parity.
This topological quantum number is similar to that of Kitaev's toric
code in that it appears to be necessary to measure a {\it global}
property of the system, using, for example, the Lieb-Schultz-Mattis
slow twist operator, in order to determine its value. Motivated by
this similarity between chiral spin liquids and toric codes, it is
natural to ask whether, or to what extent, the topologically
degenerate chiral spin liquid states on finite lattices can be viewed
as quantum error correcting codes.

A quantum error correcting code for a single qubit is a mapping of the
form, $|0\rangle \rightarrow |0_L\rangle$ and $|1\rangle \rightarrow
|1_L\rangle$, where the states $|0_L\rangle$ and $|1_L\rangle$ are
made up of several {\it physical} qubits.  If the encoded qubit is
placed in a pure state $|\Upsilon_L\rangle = \alpha |0_L\rangle +
\beta |1_L \rangle$ then the initial density matrix describing the
state is $\rho_0 = |\Upsilon_L\rangle\langle \Upsilon_L|$.  After the
physical qubits making up the encoded qubit interact with their
environment the most general effect on the density matrix is
\begin{eqnarray}
\rho_0 \rightarrow \sum_a E_a \rho_0 E^\dagger_a = \rho_E = \sum_a
E_a|\Upsilon_L\rangle\langle\Upsilon_L|E^\dagger_a ,
\end{eqnarray}
with the constraint $\sum_a E^\dagger_a E_a = 1$ where the operators
$E_a$ are referred to as error operators.  In order to be able to
return the encoded qubit to its original pure state there must exist a
recovery operation which satisfies
\begin{eqnarray}
\rho_E \rightarrow \sum_a R_a \rho_E R_a^\dagger = \rho_0=
|\Upsilon_L\rangle\langle\Upsilon_L|,
\end{eqnarray}
again with the constraint $\sum_a R^\dagger_a R_a = 1$.  The necessary
and sufficient conditions for such a recovery operation to exist are
\cite{bennett,knill}
\begin{eqnarray}
\langle 0_L | A_a^\dagger A_b | 0_L \rangle &=& \langle 1_L |
A_a^\dagger A_b | 1_L \rangle ,
\label{qec1}\\
\langle 0_L | A_a^\dagger A_b | 1_L \rangle &=& 0 ,
\label{qec2}
\end{eqnarray}
where the set of operators $\{A_a\}$ form a linear basis for the error
operators, i.e., every error operator can be expanded as $E_a = \sum_b
\lambda_{ab} A_b$.  For example, for a code capable of correcting only
single qubit error one may take the basis $\{A_a\}$ to consist of the
identity operator and all Pauli matrices acting on individual physical
qubits.

The chiral spin liquid states $\Phi_0$ and $\Phi_1$ are not
orthogonal, except in the limit $N_1/N_2 \rightarrow \infty$.
However, on any finite lattice it is possible to orthogonalize them
and use the resulting states as a quantum code where the spin-1/2
particles located at lattice sites correspond to the physical
qubits. For example,
\begin{eqnarray}
|0_L \rangle &=& |\Phi_0\rangle ,\\ | 1_L \rangle &=&
\frac{|\Phi_1\rangle - \langle \Phi_0 | \Phi_1 \rangle |\Phi_0\rangle
}{\left(1-|\langle \Phi_0 | \Phi_1 \rangle |^2\right)^{1/2}}.
\end{eqnarray}
The question to be addressed is then, to what extent do these states
satisfy the criteria (\ref{qec1}) and (\ref{qec2}) for being quantum
error correcting codes?

Because $\Phi_0$ and $\Phi_1$ are singlets it is possible to simplify
(\ref{qec1}) and (\ref{qec2}) considerably for the case of single
qubit errors by noting that an arbitrary encoded qubit
$|\Upsilon_L\rangle = \alpha |0_L \rangle + \beta |1_L \rangle$ is
also a singlet, implying that
\begin{eqnarray}
\langle \Upsilon_L | \sigma^\alpha_{{\bf r}_i} | \Upsilon_L \rangle =
0 ,
\end{eqnarray} 
for all lattice sites ${\bf r}_i$ where $\alpha = x,y,$ or $z$, thus
ensuring that (\ref{qec1}) and (\ref{qec2}) are satisfied for $A_a =
\sigma^\alpha_{{\bf r}_i}$ and $A_b = 1$.  Likewise,
\begin{eqnarray}
\langle \Upsilon_L | \sigma^\alpha_{{\bf r}_i} \sigma^\beta_{{\bf
r}_j} | \Upsilon_L \rangle = \delta_{\alpha\beta} \langle \Upsilon_L |
\sigma^z_{{\bf r}_i} \sigma^z_{{\bf r}_j} | \Upsilon_L \rangle ,
\end{eqnarray}
for all lattice sites ${\bf r}_i$ and ${\bf r}_j$.  The conditions for
a singlet state to be a quantum error correcting code capable of
correcting a single qubit error can then be shown to be equivalent to
the requirement that
\begin{eqnarray}
\langle \Upsilon_L |
\sigma^z_{{\bf r}_i} \sigma^z_{{\bf r}_j} | \Upsilon_L \rangle = C_{ij}
\label{singqec}
\end{eqnarray}
for {\it all} states $\Upsilon_L$, i.e., the spin-spin correlation
functions must be {\it identical} for any encoded state.

Because the spin-spin correlation function decays rapidly with
distance in the chiral spin liquid states \cite{kalmeyer,laughlin1}
the largest violation of (\ref{singqec}) is likely to occur for
nearest-neighbor spin correlations.  Consider these correlations for
$|\Upsilon_L\rangle = |\Phi_0\rangle$ and $|\Upsilon_L\rangle =
|\Phi_1\rangle$.  For odd values of $N_2$ there is a broken
translation symmetry in the $x$ direction, $T_x\Phi_0 = \Phi_1$ and
$T_x\Phi_1 = \Phi_0$, and, for any lattice site ${\bf r}_0$,
\begin{eqnarray}
\langle \Phi_n | \sigma^z_{{\bf r}_0} \sigma^z_{{{\bf r}_0}+b {\hat
{\bf a}}} | \Phi_n \rangle &=& \langle \Phi_n |\sigma^z_{{{\bf r}_0}+
2 b {\hat {\bf x}}} \sigma^z_{{{\bf r}_0}+ b {\hat {\bf a}}+ 2 b {\hat
{\bf x}}} | \Phi_n\rangle
\end{eqnarray}
and
\begin{eqnarray}
\langle \Phi_0 | \sigma^z_{{\bf r}_0} \sigma^z_{{\bf r}_0+b {\hat {\bf
a}}} | \Phi_0 \rangle &=& \langle \Phi_1 |\sigma^z_{{{\bf r}_0}+b
{\hat {\bf x}}} \sigma^z_{{{\bf r}_0}+b {\hat {\bf a}} + b{\hat{\bf
x}}} |\Phi_1\rangle.
\end{eqnarray}
Here, and in what follows, $\hat {\bf a} = \hat {\bf x}, \hat {\bf
y}$.  For even values of $N_2$ there is no broken translation symmetry
in the $x$ direction, $T_x \Phi_n = \Phi_n$, and
\begin{eqnarray}
\langle \Phi_n | \sigma^z_{{\bf r}_0} \sigma^z_{{{\bf r}_0+b {\hat
{\bf a}}}} | \Phi_n \rangle &=& \langle \Phi_n |\sigma^z_{{{\bf
r}_0}+b {\hat {\bf x}}} \sigma^z_{{{\bf r}_0}+b {\hat {\bf a}}+b {\hat
{\bf x}}} | \Phi_n\rangle.
\end{eqnarray}
For both even and odd values of $N_2$ there is no broken translation
symmetry in the $y$ direction, $T_y\Phi_n = \Phi_n$, and so in both
cases
\begin{eqnarray}
\langle \Phi_n | \sigma^z_{{\bf r}_0} \sigma^z_{{{\bf r}_0}+b {\hat
{\bf a}}} | \Phi_n \rangle &=& \langle \Phi_n |\sigma^z_{{{\bf r}_0}+b
{\hat {\bf y}}} \sigma^z_{{{\bf r}_0}+b {\hat{\bf a}}+b {\hat{\bf y}}}
|\Phi_n\rangle.
\end{eqnarray}
Finally, for odd values of $N_2$ the fact that the chiral spin liquid
states are symmetric under $PT$, the product of parity and
time-reversal \cite{wen1}, implies that
\begin{eqnarray}
\langle \Phi_0 | \sigma^z_{{\bf r}_0} \sigma^z_{{\bf r}_0+b {\hat {\bf
y}}} | \Phi_0 \rangle &=& \langle \Phi_1 |\sigma^z_{{\bf r}_0}
\sigma^z_{{{\bf r}_0}+b{\hat{\bf y}}}|\Phi_1\rangle.
\end{eqnarray}

Though it is not possible to compute these correlation functions
analytically, it is straightforward to compute them numerically using
standard variational Monte Carlo techniques, and the results of such
calculations for various lattice sizes are given in Table \ref{cf}.
In this Table the site ${\bf r}_0$ is taken to be the origin (${\bf
r}_0 = 0 \hat {\bf x} + 0 \hat {\bf y}$) and for each lattice size the
correlation functions $\langle \Phi_n | \sigma^z_{{{\bf r}_0}}
\sigma^z_{{{\bf r}_0}+b {\hat{\bf x}}} | \Phi_n \rangle$ and $\langle
\Phi_n | \sigma^z_{{{\bf r}_0}} \sigma^z_{{{\bf r}_0}+b {\hat{\bf y}}}
| \Phi_n \rangle$ are given for $n=0$ and 1.  Using the symmetry
properties derived above, these correlation functions can be used to
determine all nearest-neighbor spin-spin correlation functions for
$\Phi_0$ and $\Phi_1$.  (Note that for odd values of $N_2$ there is
some redundancy in the Table, since $\langle \Phi_0 | \sigma^z_{{{\bf
r}_0}} \sigma^z_{{{\bf r}_0}+b {\hat{\bf y}}} | \Phi_0 \rangle
=\langle \Phi_1 | \sigma^z_{{{\bf r}_0}} \sigma^z_{{{\bf r}_0}+b
{\hat{\bf y}}} | \Phi_1 \rangle$.)

\begin{figure}[t]
\centerline{\psfig{figure=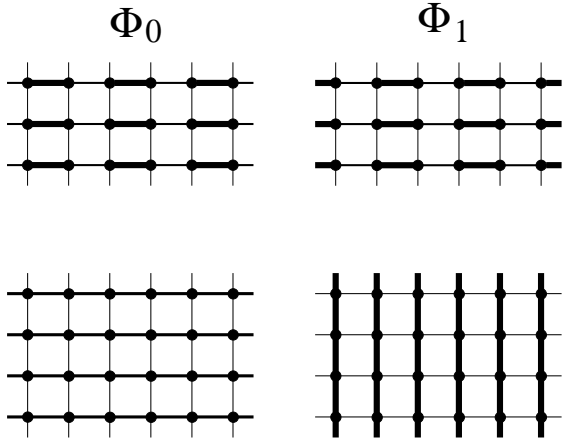,height=2.5in,angle=270}}
\vskip .15in
\caption{Patterns formed by nearest-neighbor spin-spin correlation
functions $\langle \Phi_n | \sigma_{{\bf r}_i}^z {\sigma_{{\bf
r}_j}^z} | \Phi_n \rangle$ for the topologically degenerate chiral
spin liquid states $\Phi_0$ and $\Phi_1$ on $6 \times 3$ and $6 \times
4$ lattices.  Thicker lines correspond, qualitatively, to larger
values of $-\langle \Phi_n | \sigma_{{\bf r}_i}^z \sigma_{{\bf r}_j}^z
| \Phi_n \rangle$ (thicknesses of the lines are exaggerated for
clarity).  For odd values of $N_2$ the broken translation symmetry is
observable.  For even values of $N_2$ the nearest-neighbor spin-spin
correlation functions are different in the two states.  The ability to
distinguish between $\Phi_0$ and $\Phi_1$ by measuring operators
consisting of only two Pauli matrices indicates that although the
underlying distinction between them is topological, as depicted in
\protect{Fig.~\ref{lsm}}, these states are not quantum error
correcting codes on finite lattices.}
\label{cheshire}
\end{figure}

As can be seen in Table \ref{cf}, on finite lattices the
nearest-neighbor spin-spin correlation functions are {\it not}
identical for $\Phi_0$ and $\Phi_1$, thus violating (\ref{singqec}).
Therefore, on these lattices, the topologically degenerate chiral spin
liquid states are not exact quantum error correcting codes, even for
single qubit errors.  While it is true that with increasing lattice
size the difference between correlation functions in $\Phi_0$ and
$\Phi_1$ becomes smaller, until it is no longer possible to
distinguish between them due to the statistical error bars of the
Monte Carlo simulation, given the clear violation of (\ref{singqec})
for lattices sizes as large as $8 \times 6$ it is unlikely that these
correlation functions ever become exactly equal to one another on any
finite lattice.  Rather, it is more plausible that they approach each
other exponentially as the system size, in particular $N_2$,
increases, though no proof of this has been found.

The distinction between the states $\Phi_0$ and $\Phi_1$ can be seen
clearly in the patterns formed by the values of the nearest-neighbor
spin-spin correlation functions.  These patterns are shown for $6
\times 3$ and $6 \times 4$ lattices in Fig.~\ref{cheshire}.  If the
topologically degenerate chiral spin liquid states did provide exact
quantum error correcting codes for single qubit errors then these
patterns would be identical for a given lattice size.  Figure
\ref{cheshire} together with Table \ref{cf} show clearly that despite
the fact that the underlying distinction between the states $\Phi_0$
and $\Phi_1$ is topological, as illustrated in Fig.~\ref{lsm}, on
finite lattices the difference between them can still be measured
locally using just two Pauli matrices. However, as stated above, the
nearest-neighbor spin-spin correlation functions rapidly become
effectively indistinguishable for these two states as the lattice size
increases, as do, plausibly, all the correlation functions appearing
in (\ref{singqec}).  In this sense the topologically degenerate chiral
spin liquid states on sufficiently large lattices may be viewed as
approximate quantum error correcting codes.

\section{Conclusions}

In this paper the chiral spin liquid states first introduced by
Kalmeyer and Laughlin as possible ground states for frustrated
spin-1/2 antiferromagnets have been analyzed from the point of view of
their connection to quantum error correcting codes.  Explicit wave
functions were constructed for the two topologically degenerate chiral
spin liquid states on finite periodic $N_1 \times N_2$ lattices with
$N_1$ even and it was proven that, if properly constructed, these
states are exact singlets for any such lattice.  It was also shown
that, in a sense which becomes precise when $N_1 \gg N_2$, the
property characterizing the topological degeneracy is the gap parity
--- a topological quantum number associated with the short-range
valence-bond state basis.  However, despite the fact that, like
Kitaev's toric codes, the degenerate chiral spin liquid states are
distinguished by a topological quantum number, these states are not
perfectly indistinguishable when measured with local operators, except
in the thermodynamic limit.  Thus, on finite periodic lattices, these
states do not satisfy the criteria (\ref{qec1}) and (\ref{qec2}), and
so are not exact quantum error correcting codes --- any error
correction scheme using chiral spin liquid states would not be able to
recover even a single qubit error with perfect fidelity.  However, the
distinction between these states, as measured by local operators,
rapidly becomes unobservable as the lattice size increases.
Therefore, on large enough lattices, the topologically degenerate
chiral spin liquid states may be viewed as approximate quantum error
correcting codes.

\acknowledgements 

This work was supported by the U.S.\ Department of Energy under Grant
No.\ DE-FG02-97ER45639.

\end{multicols}

\begin{table}[h]
\vskip 0.3in
\caption{Nearest-neighbor spin-spin correlation functions in the
states $\Phi_0$ and $\Phi_1$ for different lattice sizes.}
\begin{tabular}{ccccc}
Lattice Size 
& $\langle \Phi_0 | \sigma^z_{{\bf r}_0} \sigma^z_{{\bf r}_0+
b \hat  x} | \Phi_0 \rangle$ & 
$\langle \Phi_0 | \sigma^z_{{{\bf r}_0}} \sigma^z_{{{\bf r}_0}+ b \hat
y} | \Phi_0\rangle$ & $\langle \Phi_1 | \sigma^z_{{\bf r}_0}
\sigma^z_{{{\bf r}_0}+b \hat x} | \Phi_1\rangle$ & $\langle \Phi_1 |
\sigma^z_{{{\bf r}_0}} \sigma^z_{{{\bf r}_0}+ b \hat y} | \Phi_1\rangle$
\\ $N_1 \times N_2$ \\
\tableline
4 $\times$ 2 & -0.173(2) & -0.946(2) & -0.455(2) & \phantom{-}0.273(5) \\
4 $\times$ 4 & -0.247(2) & -0.246(3) & -0.230(2) & -0.376(3)\\
4 $\times$ 6 & -0.216(2) & -0.312(2) & -0.217(2) & -0.301(3)\\
4 $\times$ 8 & -0.210(2) & -0.306(3) & -0.210(2) & -0.307(3) \\
\tableline
6 $\times$ 2 & -0.176(2) & -0.944(2) & -0.467(2) & \phantom{-}0.322(5) \\
6 $\times$ 4 & -0.311(2) & -0.216(3) & -0.279(2) & -0.376(3) \\
6 $\times$ 6 & -0.298(2) & -0.300(3) & -0.302(2) & -0.281(3) \\
6 $\times$ 8 & -0.303(2) & -0.289(3) & -0.302(2) & -0.290(3) \\
\tableline
8 $\times$ 2 & -0.175(2) & -0.944(2) & -0.464(2) & \phantom{-}0.335(5)\\
8 $\times$ 4 & -0.306(2) & -0.210(3) & -0.275(2) & -0.382(3)\\
8 $\times$ 6 & -0.290(2) & -0.303(3) & -0.292(2) & -0.281(3) \\
8 $\times$ 8 & -0.291(2) & -0.291(2) & -0.290(2) & -0.293(3) \\
\tableline
\tableline
4 $\times$ 3 & -0.230(2) & -0.241(3) & -0.301(2) & -0.241(3) \\
4 $\times$ 5 & -0.229(2) & -0.301(3) & -0.221(2) & -0.301(3) \\
4 $\times$ 7 & -0.213(2) & -0.305(3) & -0.213(2) & -0.305(3) \\
4 $\times$ 9 & -0.209(2) & -0.306(3) & -0.209(2) & -0.306(3) \\
\tableline
6 $\times$ 3  & -0.334(2) & -0.239(3) & -0.257(2) & -0.239(3)  \\
6 $\times$ 5  & -0.280(2) & -0.290(3) & -0.292(2) & -0.290(3) \\
6 $\times$ 7  & -0.283(2) & -0.294(3) & -0.281(2) & -0.294(3) \\
6 $\times$ 9  & -0.281(2) & -0.293(3) & -0.281(2) & -0.293(3) \\
\tableline
8 $\times$ 3 & -0.258(2) & -0.239(3) & -0.336(2) & -0.239(3) \\
8 $\times$ 5 & -0.298(2) & -0.290(3) & -0.293(2) & -0.290(3) \\
8 $\times$ 7 & -0.291(2) & -0.290(3) & -0.292(2) & -0.290(3) \\
8 $\times$ 9 & -0.290(2) & -0.291(3) & -0.290(2) & -0.291(3) 
\end{tabular}
\label{cf}
\end{table}

\end{document}